\documentclass[rmp,aps]{revtex4-1}
\pdfoutput=1
\usepackage{latexsym}
\usepackage[pdftex]{graphicx}
\usepackage{amsmath}
\usepackage{natbib}

\begin{document}
\title{Bose-Hubbard Hamiltonian: Quantum Chaos approach}
\author{Andrey R. Kolovsky}
\affiliation{Kirensky Institute of Physics, 660036 Krasnoyarsk Russia}
\affiliation{Siberian Federal University, 660041 Krasnoyarsk Russia}
\date{\today}

\begin{abstract}
We discuss applications of the theory of Quantum Chaos to one of the paradigm models of many-body quantum physics -- the Bose-Hubbard model, which describes, in particular, interacting ultracold Bose atoms in an optical lattice. After preliminary, pure quantum analysis of the system we introduce the classical counterpart of the Bose-Hubbard model and the governing semiclassical equations of motion. We analyze these equations for the problem of Bloch oscillations of cold atoms where a number of experimental results are available. The review is written for non-experts and can be viewed as an introduction to the field.
\end{abstract}
\maketitle

\tableofcontents

\section{Introduction}

The Bose-Hubbard (BH) model, introduced by Gersch and Knollman in 1963 \cite{Gers63}, became a hot topic in physics after the theoretical work by Jaksch {\em et. al.} of 1998 \cite{Jaks98}, where it was noticed that this model can be realized with the help of cold atoms loaded into an optical lattice, and, especially, after the fundamental experiment by Greiner  {\em et. al.} of 2002 \cite{Grei02}, where the authors demonstrated quantum phase transition from the super-fluid state to the Mott insulator for rubidium atoms in a cubic optical lattice. Ever since the BH model is discussed almost exclusively in the context of cold atoms physics \cite{Jaks05}.

In the past decade many different phenomena of the cold atoms physics, which are described by the Bose-Hubbard and Bose-Hubbard like Hamiltonians, were studied in great detail. Naturally, these studies contributed to our understanding of the properties of the model, which is currently considered as paradigm model of many-body physics that can be tested experimentally. Still, we are far from complete understanding and every novel analytical approach or laboratory experiment adds something new to our knowledge of the BH model.

In this tutorial review we describe a new approach to the BH model which is based on the ideas of Quantum Chaos -- a modern theory which deals with quantum non-integrable systems \cite{Gian91,Stoe99}. A particular feature of this approach is the extensive use of classical mechanics. This might come as a surprise because the BH model is usually considered as a genuine quantum system, with no classical counterpart. However, as it will be explained in the review, there is an analogy between the quantum and classical descriptions of a single-particle system, and the microscopic and mean-field descriptions of a many-body system. This analogy helps us to better understand the BH model, especially, when it concerns excited states of the system.

The review consists of three parts. In Sec.~\ref{secA} we consider the BH model as a generic complex quantum system, without appealing to the classical mechanics.  Following the main idea of Quantum Chaos that the energy spectrum of a complex system should have common features with the spectrum of random matrices, we perform statistical analysis of eigenvalues and eigenfunctions of the BH Hamiltonian and compare the result with RMT (Random Matrix Theory) predictions. Section \ref{secB} begins with discussing of the semiclassical limit, where we follow the method of the truncated Husimi function. The power of this semiclassical method is demonstrated in Sec.~\ref{F}, where we derive the Bogoliubov spectrum by using  semiclassical quantization, and in Sec.~\ref{secC}, where we consider the problem of Bloch oscillations (BOs) of cold atoms in tilted 1D optical lattices. It is shown, in particular, that one can reproduce quantum dynamics of the system by solving classical equations.

\section{Energy spectrum of the Bose-Hubbard model}
\label{secA}

\subsection{The Bose-Hubbard Hamiltonian}
\label{A}

Having in mind cold Bose atoms in a 1D optical lattice
\footnote{One-dimensional optical lattice or, more exactly, an array of independent 1D lattices is created by using two strong standing laser waves in the $x$ and $y$ directions, which create the so-called quantum tubes, and one weak standing wave in the $z$ direction, which periodically modulates the quantum tubes.}
the BH Hamiltonian reads
\begin{equation}
\label{A1a}
\widehat{H}_0=-\frac{J}{2} \sum_{l=1}^L \left( \hat{a}^\dag_{l+1}\hat{a}_l +h.c.\right)
  +\frac{U}{2}\sum_{l=1}^L \hat{n}_l(\hat{n}_l-1) \;.
\end{equation}
In Eq.~(\ref{A1a}) the index $l$ labels the lattice sites (wells of the optical potential), $\hat{a}_l$ and $\hat{a}^\dagger_l$ are the bosonic annihilation and creation operators, 
\begin{displaymath}
[\hat{a}_l,\hat{a}^\dagger_{l'}]=\hbar\delta_{l,{l'}} \;,
\end{displaymath}
$\hat{n}_l=\hat{a}_l^\dagger\hat{a}_l$ is the number operator, $J$ the hopping matrix element (the rate of inter-well tunneling), and $U$ the microscopic interaction constant (the energy  paid by two atoms sharing the same well). The constant $U$ is mainly determined by the $s$-wave scattering length $a_s$ for neutral atoms,
\begin{displaymath}
U=\frac{4\pi a_s \hbar^2}{M}\int |\phi_l({\bf r})|^4 d^3{\bf r} \;,
\end{displaymath}
and the constant $J$ by the lattice depth $V_0$,
\begin{displaymath}
J=\int \phi_{l+1}(z) \widehat{H}_s \phi_l(z) dz \;,\quad
\widehat{H}_s=-\frac{\hbar^2}{M}\frac{\partial^2}{\partial z^2} + V_0\cos^2(k_L  z)
\end{displaymath}
(here $\phi_l(z)$ are the Wannier functions localized  at $l$th well of the optical potential and $k_L$ is the laser wave vector). Both the scattering length and the depth $V_0$ can be varied in large intervals, which allows practically arbitrary ratio $U/J$.
\begin{figure}[b]
\includegraphics[width=10cm, clip]{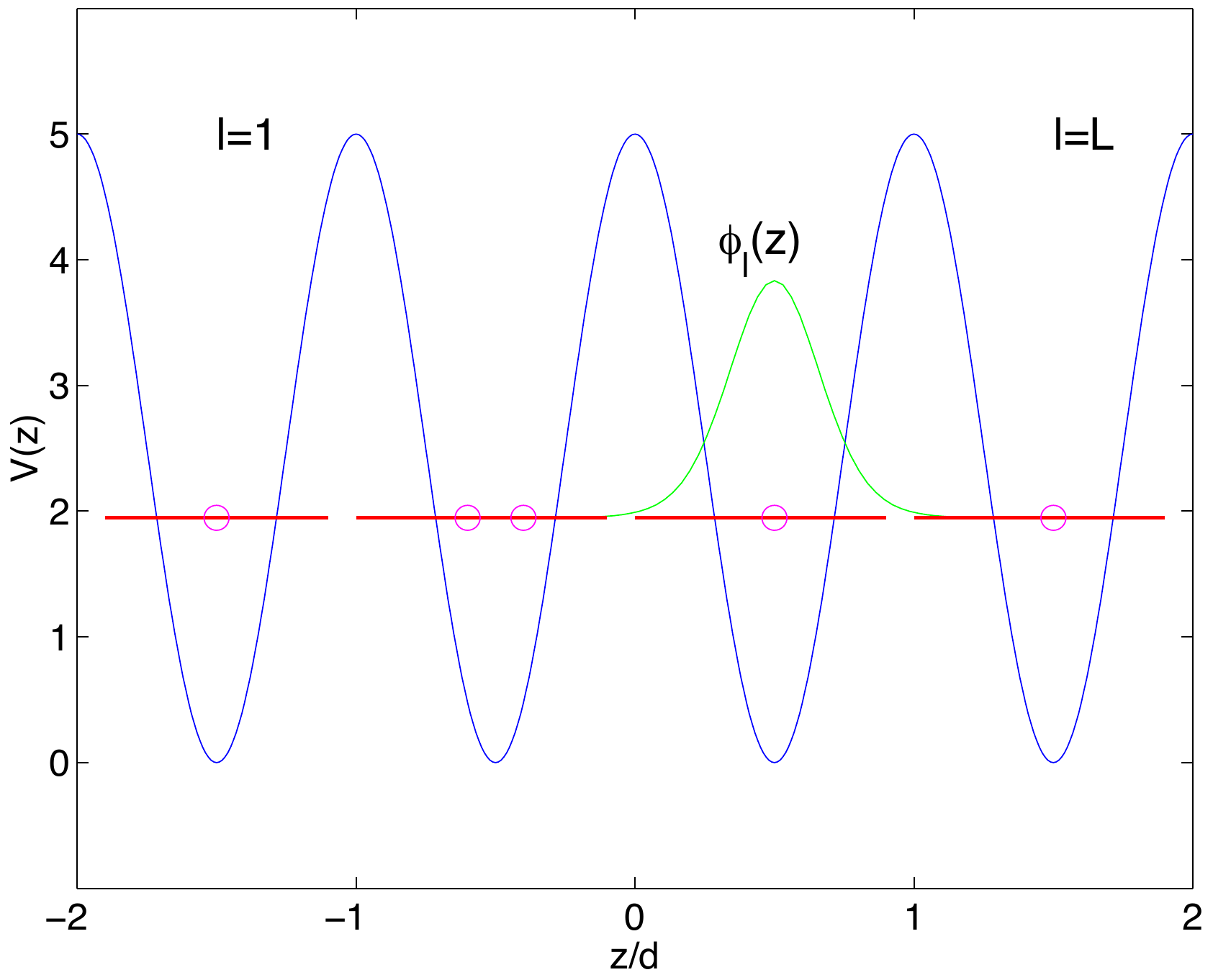}
\caption{Pictorial presentation of cold atoms (open circles) in an optical lattice. The thin green line shows one of the Wannier function $\phi_l(z)$.}
\label{fig0}
\end{figure}

A remark concerning boundary conditions is in turn. In a laboratory experiment the default boundary condition is residual harmonic confinement due to finite widths of the laser beams. In the theory, however, one usually considers the periodic boundary condition, where $(L+1)$th site of the lattice is identified with the first site, i.e., $\hat{a}^\dagger_{L+1}=\hat{a}^\dagger_1$. 
\footnote{We mention, in passing, that few-site BH model with periodic boundary condition can be realized experimentally by using non-trivial Gaussian beams \cite{Amic05}. This setup, however, excludes consideration of the limit $L\rightarrow\infty$.}
Through the paper we shall assume the later case. Notice that the periodic boundary condition implies conservation of the total quasimomentum. This can be seen by rewriting the Hamiltonian (\ref{A1a}) in terms of the operators $\hat{b}_k$ and $\hat{b}^\dagger$,
\begin{equation}
\label{A3}
\hat{b}_k=\frac{1}{\sqrt{L}} \sum_l \exp\left(i\frac{2\pi k}{L}l\right) \hat{a}_l \;, \quad 
\hat{b}^\dagger_k=\left(\hat{b}_k\right)^\dagger \;.
\end{equation}
Operators (\ref{A3}) annihilate or create an atom in the Bloch state with the quasimomentum $\kappa=2\pi k/L$. Using the transformation (\ref{A3}) we have
\begin{equation}
\label{A1b}
\widehat{H}_0=-J\sum_k \cos\left(\frac{2\pi k}{L}\right)\hat{b}_k^\dag\hat{b}_k
  +\frac{U}{2L}\sum_{k_1,k_2,k_3,k_4}
  \hat{b}_{k_1}^\dag\hat{b}_{k_2}^\dag\hat{b}_{k_3}\hat{b}_{k_4}
  \tilde{\delta}(k_1+k_2-k_3-k_4) \;,
\end{equation}
where $\tilde{\delta}$ is the periodic $\delta$-function, i.e., $\tilde{\delta}(k)$ equals unity if $k$ is a multiple of $L$ and zero otherwise. The presence of  the $\delta$-function in the interaction term insures that the total quasimomentum is conserved.
\begin{figure}
\includegraphics[width=6.5cm, clip]{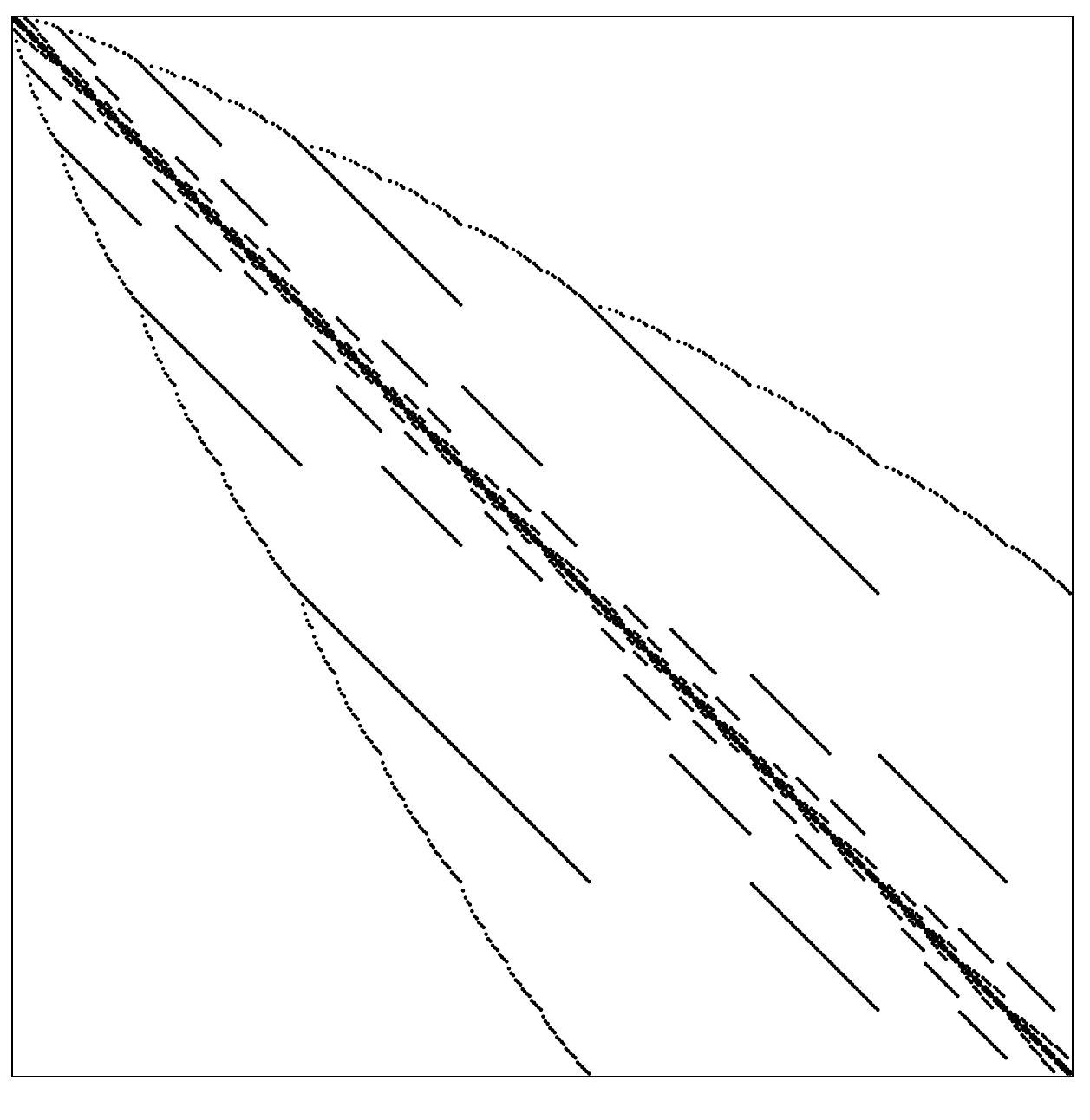}
\hspace{2cm}
\includegraphics[width=6.5cm, clip]{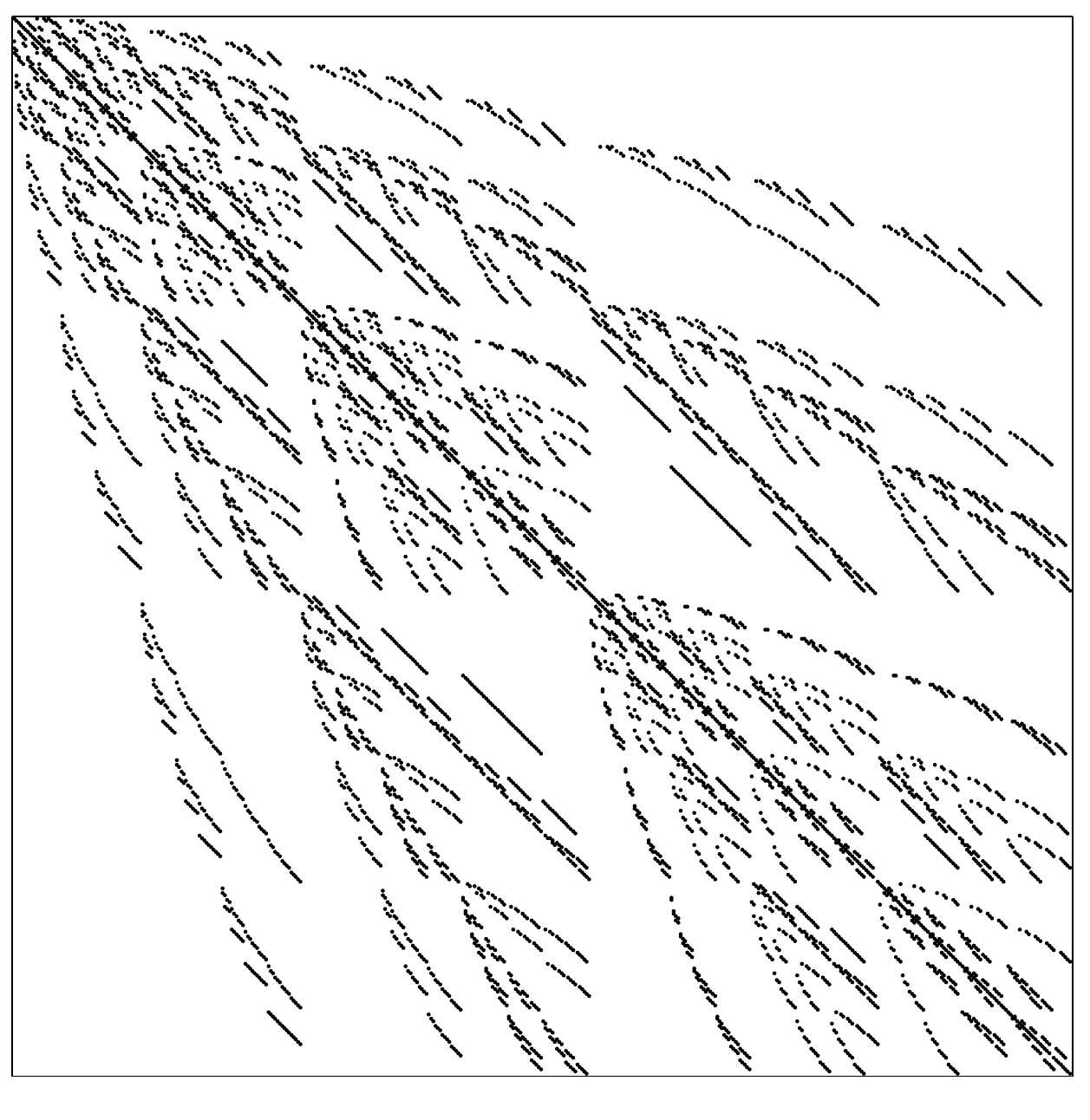}
\caption{Example of the Hamiltonian matrix for $N=L=5$. Shown are nonzero matrix elements of the Hamiltonians (\ref{A1a}) and (\ref{A1b}) in the Fock basis (\ref{A3a}) and (\ref{A3c}), respectively.}
\label{fig1}
\end{figure}

Finally, we discuss the Hilbert space of the BH system. It is spanned by the Fock states
\begin{equation}
\label{A3a}
|{\bf n}\rangle=|n_1,n_2,\ldots,n_L\rangle \;,\quad \sum_l n_l=N \;,
\end{equation}
where  $N$ is the total number of atoms. The dimension of the Hilbert space is
\begin{displaymath}
{\cal N}=\frac{(N+L-1)!}{N!(L-1)!} \;.
\end{displaymath}
In the coordinate representation the basis state (\ref{A3a}) is given by the symmetrized product of $N$ Wannier functions $\phi_l(z)$. Correspondently, if we consider the basis state of the Hamiltonian (\ref{A1b}), 
\begin{equation}
\label{A3c}
|{\bf n}\rangle=|n_1,n_2,\ldots,n_L\rangle \;,\quad \sum_k n_k=N \;,
\end{equation}
it is given by the symmetrized product of $N$ Bloch waves $\Phi_k(z)=L^{-1/2}\sum_l \exp(i2\pi kl/L) \phi_l(z)$. The total quasimomentum of the state (\ref{A3c}) is calculated as
\begin{equation}
\label{A3d}
\kappa=\frac{2\pi}{L}{\rm mod}_L\left(\sum_k k n_k\right) 
\end{equation}
and can take one of $L$ values. Knowing the action of bosonic operators on a given Fock state, 
\begin{displaymath}
\hat{a}_l |\ldots,n_l,\ldots \rangle=\sqrt{n_l}|\ldots,n_l-1,\ldots \rangle \;,\quad
\hat{a}^\dagger_l |\ldots,n_l,\ldots \rangle=\sqrt{n_l+1}|\ldots,n_l+1,\ldots \rangle \;,
\end{displaymath}
(for operators $\hat{b}_k$ and $\hat{b}_k^\dagger$ we have similar equations) we calculate the Hamiltonian matrix of the size ${\cal N}\times{\cal N}$. The explicit form of this matrix depends on particular ordering of the basis states. The main point, however, is that this matrix appears to be very sparse, see Fig.~(\ref{fig1}).

\subsection{Statistical analysis of the energy spectrum}
\label{B}
\begin{figure}
\includegraphics[width=11cm, clip]{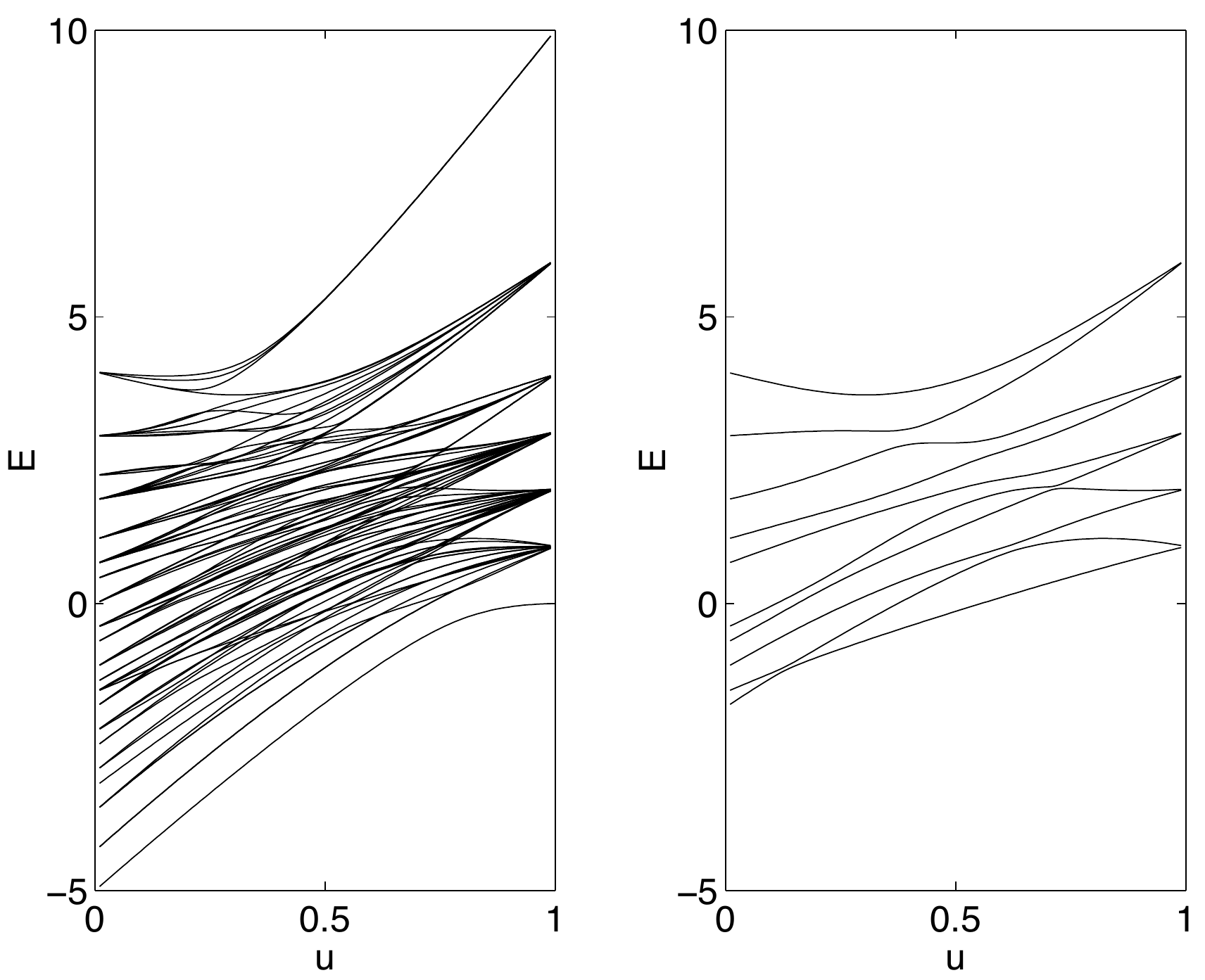}
\caption{Energy levels of the BH system ($N=L=5$) as functions of the parameter $u$ where $J=1-u$ and $U=u$ .  The left panel shows the whole spectrum, the right panel is independent subset of odd states with zero total quasimomentum. The figure is borrowed from Ref.~\cite{66}.}
\label{fig2}
\end{figure}

First we give a numerical evidence that the BH model belongs to the class of quantum non-integrable systems.  Figure \ref{fig2} shows the energy spectrum of the system for $L=N=5$ which is parametrized by the parameter $u$, where $J=1-u$ and $U=u$. This spectrum can be decomposed into $L$ independent spectra associated with different values of the total quasimomentum. In the Bloch representation (\ref{A1b}) one finds independent spectra by considering different subsets of the Hilbert space labeled by the quasimomentum (\ref{A3d}).  Alternatively, we can find these spectra by calculating the matrix of the Hamiltonian (\ref{A1a}) in the translationally invariant  basis 
\begin{displaymath}
|{\bf m},\kappa\rangle=(1/\sqrt{L})\sum_{l=1}^L \exp(i\kappa l) \widehat{S}^l |n_1,n_2,\ldots,n_L\rangle \;,
\end{displaymath}
where $\kappa$ is the total quasimomentum and $\widehat{S}$ is the cyclic permutation operator:  $\widehat{S} |n_1,n_2,\ldots,n_L\rangle = |n_2,n_3,\ldots,n_1\rangle$. The spectrum associated with $\kappa=0$ can be decomposed further according to the reflection (odd-even) symmetry. At this stage the decomposition is complete. As an example, the right panel in  Fig.~\ref{fig2} shows one of the independent spectra. It is seen that the energy levels exhibit avoided crossings as they approach each other. This proves that the BH model is not integrable.
\footnote{Exclusions are the cases $U=0$ or $J=0$ and the case of two-site BH model ($L=2$), where the spectrum can be found analytically for arbitrary $U$ and $J$.}
Moreover, it appears to be a chaotic system in the sense of Quantum Chaos. To prove the last statement one should perform statistical analysis of the spectrum, where the simplest test is the distribution of normalized distances between the nearest levels.

\begin{figure}
\includegraphics[width=11cm, clip]{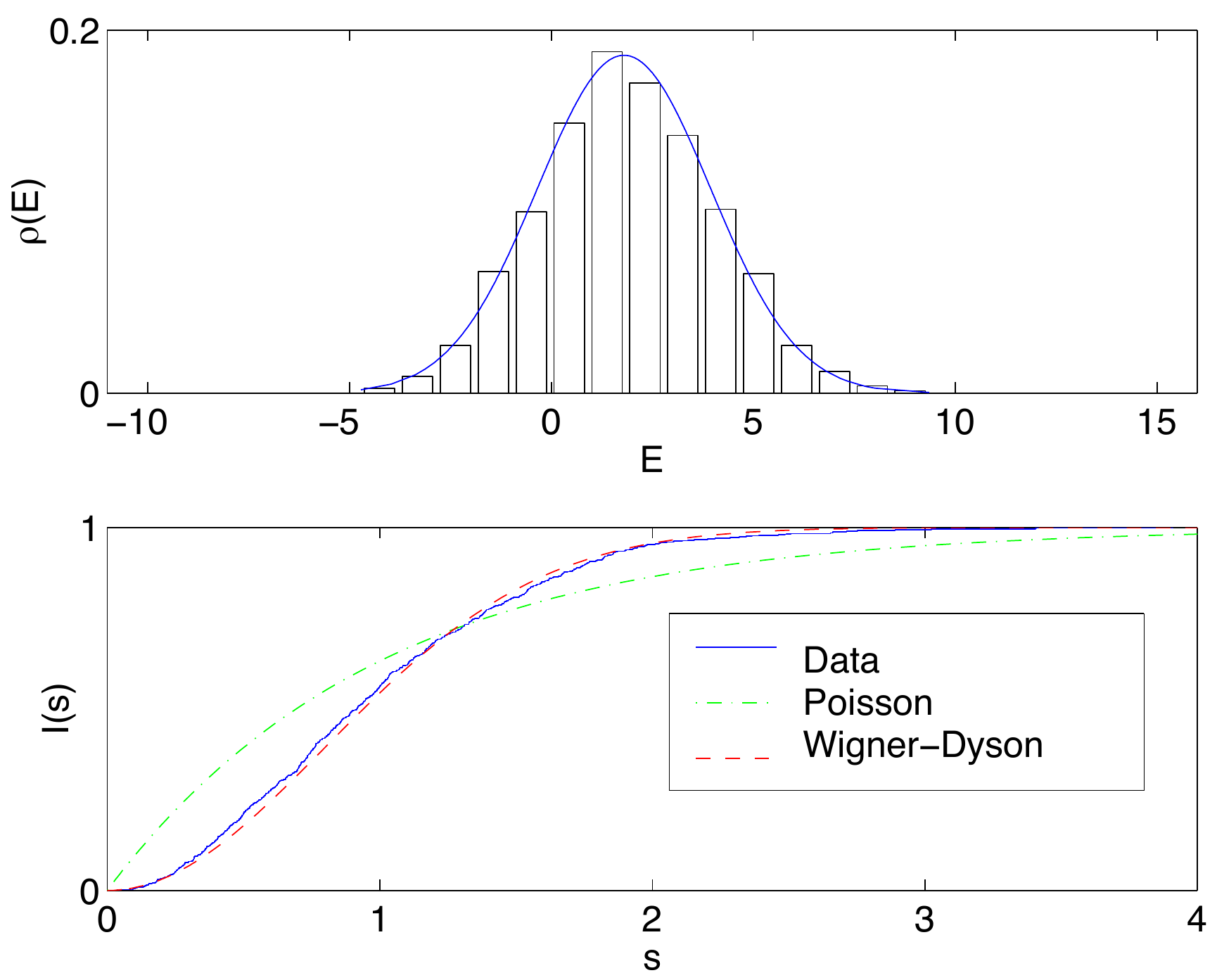}
\caption{Upper panel: The mean density of states, histogram, approximated by Eq.~(\ref{B2}), solid line. Lower panel: Integrated distribution $I(s)=\int_{-\infty}^s P(s') ds'$ for one of independent spectra as compared to the Poisson and Wigner-Dyson distributions. The system parameters are $L=N=8$ and $u=0.3$.}  
\label{fig3}
\end{figure}

To normalize the distances or, using the Quantum Chaos terminology, to unfold the spectrum we need to know the mean density of states $\rho(E)$. An estimate for $\rho(E)$ can be obtained as follows. Let us set for the moment $U=0$. Then the spectrum is known analytically [see Eq.~(\ref{A1b})]: 
\begin{displaymath}
E=-J\sum_k \cos\left(\frac{2\pi k}{L}\right)n_k \;.
\end{displaymath}
Here $n_k$ are integer numbers and, if $n_k>1$, this should be viewed as the sum of $n_k$ equal terms. Thus the energy level $E$ is given by a sum of $N$ real numbers, each restricted by modulus by $J$. Considering the thermodynamic limit $N,L\rightarrow \infty$ and using the central limit theorem we conclude that levels $E$ are distributed according to the normal law with the variance $\sigma^2\sim JN$. The normal distribution proves to be a good approximation also for nonzero $U<J$.  In this case the Gaussian is shifted as the whole by the mean interaction energy:
\begin{equation}
\label{B2}
\rho(E)\sim \exp\left[-\frac{(E-E_{int})^2}{2\sigma^2}\right] \;,\quad \sigma^2\sim JN \;,\quad E_{int}\sim \frac{UN^2}{L} \;.
\end{equation}
Validity of the approximation (\ref{B2}) is illustrated in the upper panel in Fig.~\ref{fig3}. A more thorough analysis reveals deviations of the actual density of states from Eq.~(\ref{B2}), especially at the tails of the distribution. However, for the sake of statistical analysis of the spectrum, where the main contribution comes from the central part of the spectrum, it is quite satisfactory.

Having the mean density of state obtained we introduce the normalized distance between the nearest energy levels,
\begin{equation}
\label{B3}
s=(E_{n+1}-E_{n})\rho(E_n) \;,
\end{equation}
calculate the distribution function $P(s)$ for the distance (\ref{B3}), and compare it with the Wigner-Dyson distributions for random matrices. This comparison reveals remarkable agreement with $P(s)$ for the Gaussian Orthogonal Ensemble (GOE) of random matrices,
\footnote{These are real symmetric matrices with random entries according to the normal low. More exactly, probability density to meet matrix $H$ in the ensemble is given by ${\cal P}(H)\sim\exp[-A{\rm Tr}( H^2)]$ where $A$ is the normalization constant. A common choice is $A=\pi^2/2{\cal N}$ where ${\cal N}$ is the matrix size. Then the mean density of states of the GOE matrix is given by $\rho(E)=\sqrt{1-(\pi E/2{\cal N})^2}$.}
%
\begin{equation}
\label{B4}
P(s)=\frac{\pi}{2}s\exp\left(-\frac{\pi}{4}s^2\right) \;,
\end{equation}
see lower panel in Fig.~\ref{fig3}. Thus the system (\ref{A1a}) is chaotic in the sense of Quantum Chaos.  It will be shown later on in Sec.~\ref{secB} that the BH model is a chaotic system also in the sense of Classical Chaos. This explains the amazing fact that the sparse matrix shown in Fig.~\ref{fig1} has similar properties as a fully random matrix.

\subsection{Transition to chaos}
\label{C}

The transition to chaos in the BH model takes place as a transition over the parameter $U/J$, which alone defines the properties of the system. To study this transition we can avoid complex procedure of the spectrum decomposition by introducing a weak on-site disorder,
\begin{equation}
\label{C1}
\widehat{H}=\widehat{H}_0 + \sum_{l=1}^L \epsilon_l \hat{n}_l \;, \quad |\epsilon_l |\le \epsilon \;.
\end{equation}
If the disorder is weak enough, so that the Anderson localization length is much larger than the system size $L$, the main effect of the disorder is destruction of the global symmetries (which are the total quasimomentum and odd-even symmetry). Thus we may study the spectrum statistics without preliminary decomposition of the spectrum. 
\begin{figure}
\includegraphics[width=11cm, clip]{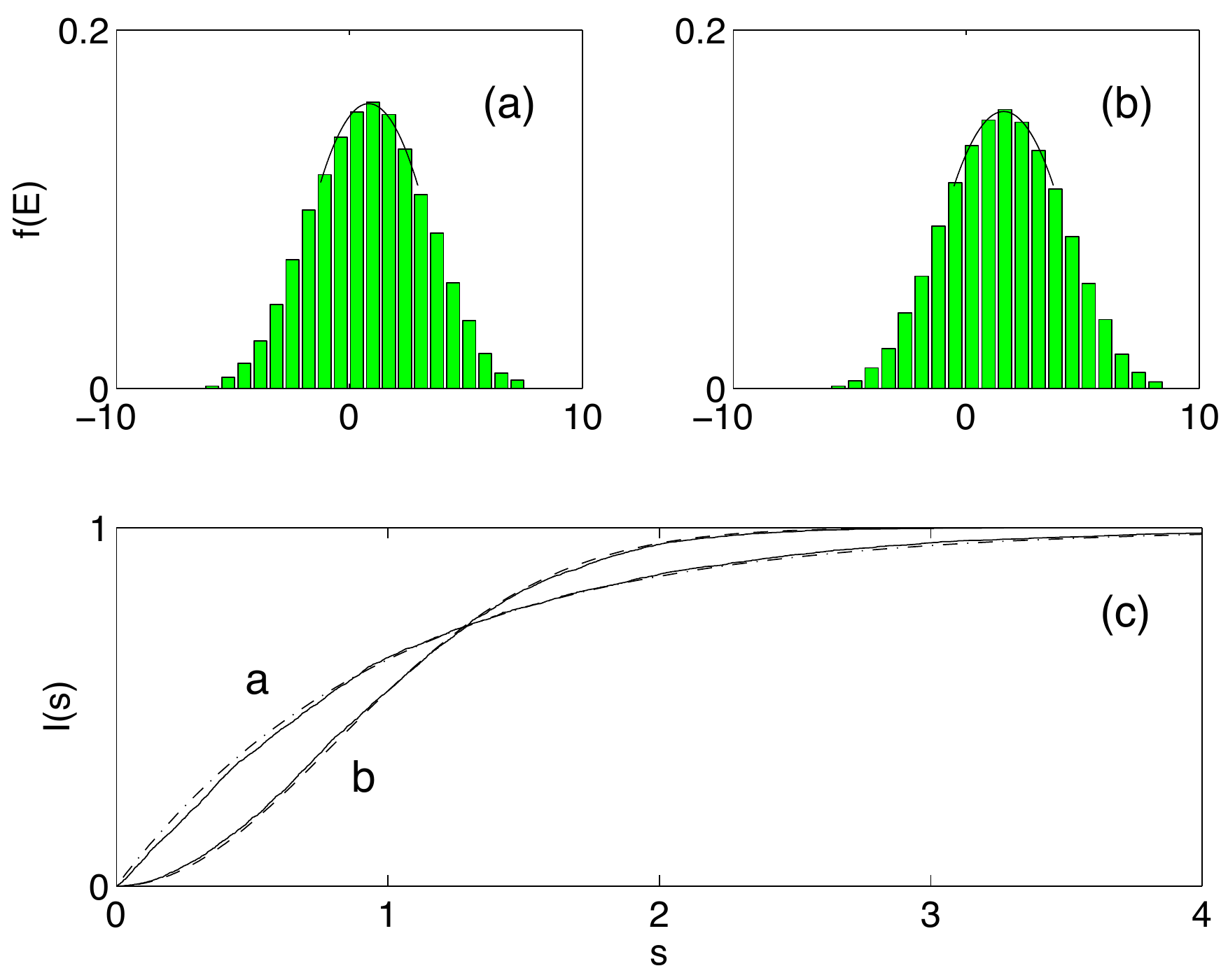}
\caption{ Panels (a) and (b): Density of states for $N=7$, $L=9$, $J=1$, $\epsilon=0.1$, and the interaction constant $U=0.02$ and $U=0.2$, respectively. Panel (c): Integrated level spacing distributions for the central part of the spectrum. The figure is borrowed from Ref.~\cite{70}.}
\label{fig4}
\end{figure}

Figure \ref{fig4} illustrates transition to chaos in the system (\ref{C1}) as we increase the interaction constant $U$. One clearly observes the change from the Poisson statistics, 
\begin{equation}
\label{C2}
P(s)=\exp(-s) \;,
\end{equation}
which is typical for integrable systems, to the Wigner-Dyson statistics (\ref{B4}), which is a hallmark of quantum chaotic systems \cite{Stoe99}.

\begin{figure}
\includegraphics[width=9cm, clip]{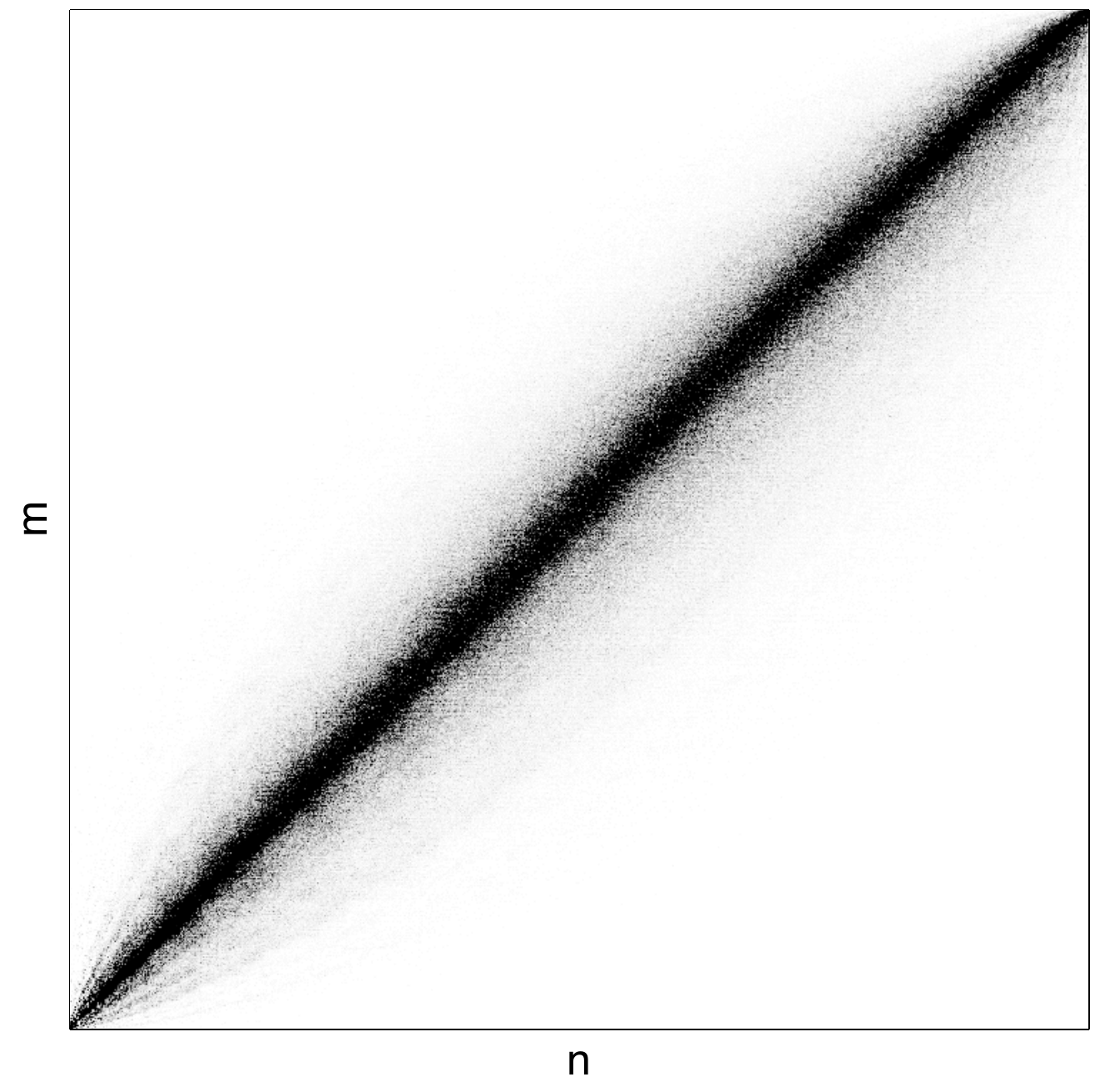}
\caption{Gray-scale image of the matrix (\ref{C3}). The system parameters are the same as in Fig.~\ref{fig4}.}
\label{fig5}
\end{figure}

Another indication of the transition to chaos comes from analysis of the eigenstates $|\Psi_n\rangle$. Note that the only quantum number of the eigenstate $|\Psi_n\rangle$ is its energy $E_n$ and we assume that the states are ordered according to their energies, i.e., $E_n>E_m$ if $n>m$. A useful characteristic of eigenstates is the matrix 
\begin{equation}
\label{C3}
R(m,n)=|\langle \Psi_m(U')|\Psi_n(U)\rangle|^2 \;,
\end{equation}
where $U$ and $U'$ are two different values of our control parameter. 
\footnote{This matrix is closely related to the so-called local density of states, $R(m,E)=\sum_n R(m,n)\delta(E-E_n)$, which has a number of important physical applications, see Ref.~\cite{79}, for example.}
Clearly, $R(n,m)$ is the identity matrix if $U=U'$. However, if  $U'$ deviates from $U$ it become a banded matrix, see Fig.~\ref{fig5}. The crucial point is that for chaotic systems $R(n,m)$ obeys the universal distribution 
\begin{equation}
\label{C4}
\bar{R}(n-m)=\frac{\Gamma/2\pi}{(n- m)^2+\Gamma^2/4}  \;,
\end{equation}
known as the Breit-Wigner equation \cite{Fyod96}. In this equation the parameter $\Gamma$  is a function of the difference $U'-U$ and the bar denotes an average over several eigenstates. (Without this averaging procedure, the matrix elements $R(n,m)$ show strong fluctuations, see Fig.~\ref{fig5}.) 

\begin{figure}
\includegraphics[width=11cm, clip]{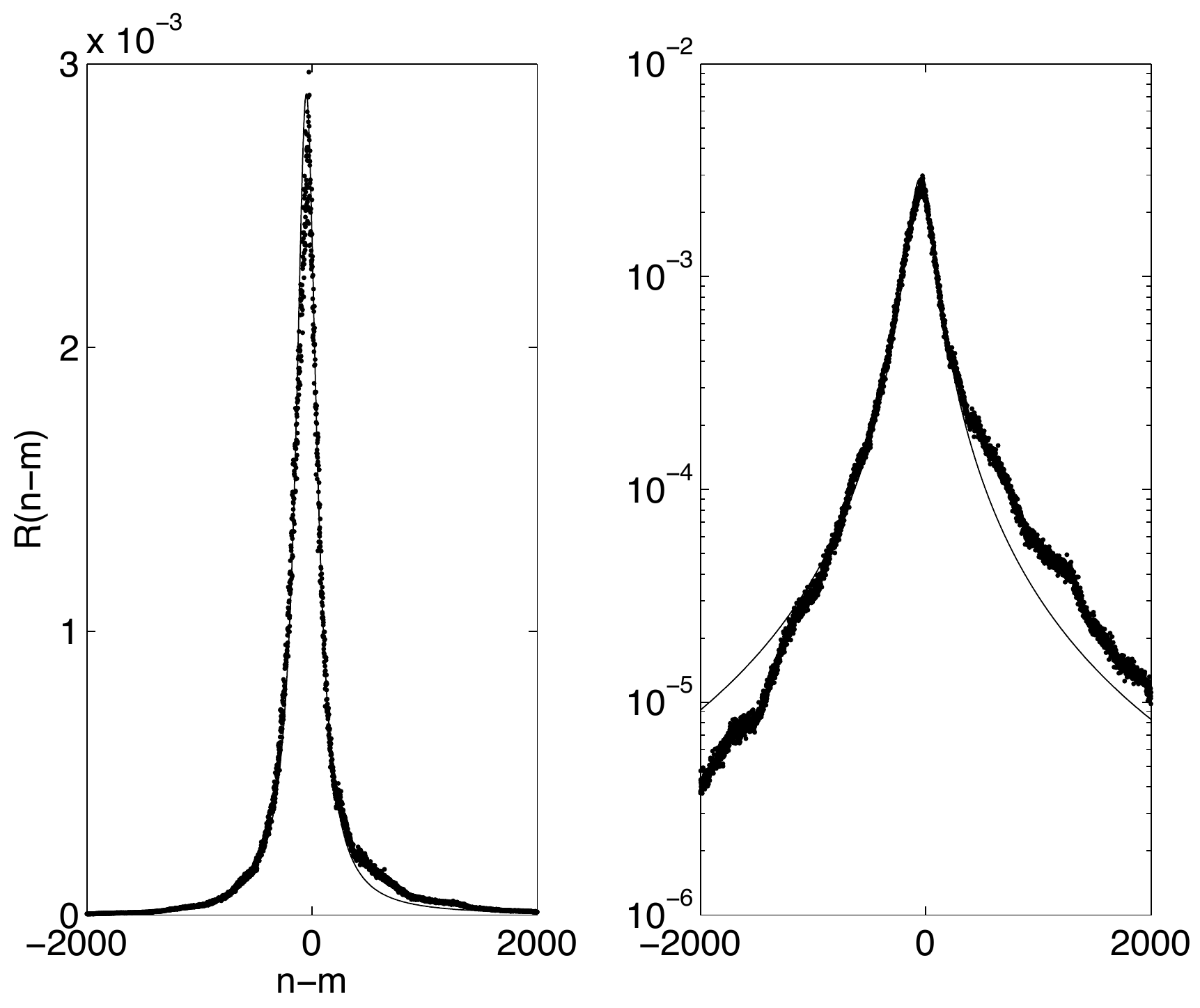}
\caption{ Mean values of the matrix elements across the main diagonal in the central part of the matrix on the linear (left panel) and logarithmic (right panel) scales. The solid line is the best fit by the Breit-Wigner formula.}
\label{fig6}
\end{figure}

The Breit-Wigner distribution (\ref{C4}) is illustrated in Fig.~\ref{fig6}. In this figure the dots are numerical data and the solid line is Eq.~(\ref{C4}). A reasonable agreement is noticed. We mention that numerical data in Fig.~\ref{fig6} are collected from the central part of the spectrum which is marked by the solid lines in Fig.~\ref{fig4}(a,b). For the low- and high-energy eigenstates one finds essential deviations from Eq.~(\ref{C4}). This observation tells that these states are not chaotic or, at least, not fully chaotic. We come back to this point in Sec.~\ref{E}. 

To conclude this section we briefly discuss a transition from GOE to GUE (Gaussian Unitary Ensemble)
\footnote{This is the ensemble of hermitian random matrices with the probability density ${\cal P}(H)\sim \exp[-A{\rm Tr}(H^\dagger H)]$.}
spectrum statistics in the BH model. The GOE statistics is typical for quantum systems with time-reversal symmetry,  which is obviously the case of the Hamiltonian (\ref{A1a}). One can break this symmetry by introducing the complex hopping matrix element: 
\begin{equation}
\label{C5}
 \widehat{H}=-\frac{J}{2} \sum_{l=1}^L \left( \hat{a}^\dag_{l+1}\hat{a}_l e^{i\theta}+h.c.\right)
  +\frac{U}{2}\sum_{l=1}^L \hat{n}_l(\hat{n}_l-1) \;.
\end{equation}
The Hamiltonian (\ref{C5}) appears in the problem of atomic Bloch oscillations and can be experimentally realized by shaking the lattice with a proper frequency \cite{84}. The effect of non-zero phase $\theta$ on dynamics of cold atoms was studied, for example, in Ref.~\cite{Hall10}. In this section our prime interest is the energy spectrum, where one observes a change of the GOE statistics (\ref{B4}) to the GUE statistics, 
\begin{displaymath}
P(s)=\frac{32}{\pi^2}s^2\exp\left(-\frac{4}{\pi}s^2\right) \;,
\end{displaymath}
as $\theta$ deviates from zero. This, however, does not affect the main result of this section that the BH model is a quantum chaotic system.

%
\section{Semiclassical quantization of the Bose-Hubbard  model}
\label{secB}

In the theory of Quantum Chaos analysis of a quantum system is usually preceded by the analysis of its classical counterpart. In this review we reverted this sequence for the reason that the quantum analysis is actually simpler. Now we come to the classical consideration, where the first step is to identify the classical counterpart of the BH model.

\subsection{Semiclassical limit}
\label{D}

There are several ways to introduce classical counterpart of the quantum system (\ref{A1a}). We shall follow the approach based on the notion of the Husimi function. 
\footnote{A similar approach is based on the notion of the Wigner function \cite{Stee98,Sina02,Polk09}. The Husimi function, however, has an advantage that it is positively defined.}
Below we shall introduce the effective Planck constant $\hbar_{eff}$ which is inverse proportional to number of atoms, $\hbar_{eff}=1/N$. This constant should not be mismatched with the fundamental Planck constant $\hbar$ which we set to unity from now on.

Given $|\Psi(t)\rangle$ to be the many-body wave function of the quantum Hamiltonian, the Husimi function is defined as
\begin{equation}
\label{D1}
f({\bf a},t)=|\langle {\bf a} |\Psi(t)\rangle |^2 \;,
\end{equation}
where $|{\bf a}\rangle$ are the so-called coherent $SU(L)$ states \cite{Pere86},
\begin{displaymath}
 |{\bf a}\rangle= \frac{1}{\sqrt{N!}}\left(\sum_{l=1}^L a_l \hat{a}^\dagger_l \right)^N |vac\rangle \;.
\end{displaymath}
Note that the Husimi function (\ref{D1}) is a function of $L$ complex amplitude $a_l$ and time. In terms of the Husimi function (\ref{D1}) the Schr\"odinger equation for the wave function $|\Psi(t)\rangle$ takes the form 
\begin{equation}
\label{D3}
\frac{\partial f}{\partial t}=\{H,f\}+O\left(\frac{1}{N}\right) \;,
\end{equation}
where $\{\ldots,\ldots\}$ denotes the Poisson brackets, the $c$-number Hamiltonian $H_0$ reads
\begin{equation}
\label{D4}
H_0=-\frac{J}{2}\sum_{l=1}^L (a^*_{l+1}a_l + c.c.) + \frac{g}{2}\sum_{l=1}^L |a_l|^4  \;, \quad g=\frac{UN}{L} \;,
\end{equation}
and we refer the reader to the work \cite{Trim08a} for explicit form of the terms which are inverse proportional to $N$. The constant $g$ in the classical Hamiltonian (\ref{D4}) is called the macroscopic interaction constant, to distinguish it from the microscopic interaction constant $U$.

We remark that, formally, the Hamiltonian (\ref{D4}) follows from the microscopic Hamiltonian (\ref{A1a}) by using the `quantization rules'  $a_l \leftrightarrow \hat{a}_l/\sqrt{\bar{n}}$, $a^*_l\leftrightarrow \hat{a}^\dagger_l/\sqrt{\bar{n}}$, and $H \leftrightarrow \widehat{H}/\bar{n}$, where $\bar{n}=N/L$ is the filling factor (the mean number of atoms per lattice site). The approach of the Husimi function (as well as the approach of the Wigner function) provides a justification of these quantization rules.

Let us now consider the mean-field limit $N\rightarrow\infty$, $U\rightarrow0$ while $g=const$. In this limit the terms proportional to $1/N$ vanish and the Husimi function reduces to the multidimensional $\delta$-function,
\begin{displaymath}
f({\bf a},t)=\prod_{l=1}^L\delta(a_l-a_l(t)) \;,
\end{displaymath}
where $a_l(t)$ satisfies the Hamilton equations of motion,
\begin{equation}
\label{D6}
i\frac{d}{dt} a_l=\frac{\partial H_0}{\partial a^*_l} = -\frac{J}{2}\left(a_{l+1} +a_{l-1}\right) + g|a_l|^2 a_l \;.
\end{equation}
Eq.~(\ref{D6}) is known in the physical literature as the Discrete Nonlinear Schr\"odinger Equation (DNLSE) and can be viewed as a discrete analogue of the Gross-Pitaevskii equation for a Bose-Einstein condensate \cite{Pita03}. We mention that DNLSE also appears in problems of Nonlinear Optics and molecular vibrations, where it has been studied for decades \cite{Eilb85}.

A comment is due on the terminology. In the title of this section we used `semiclassical limit' instead of the `mean-field limit'. The reason is that Eq.~(\ref{D3}) formally coincides with equation on the Husimi function of a single-particle system if one identifies $1/N$ with the Planck constant. Thus one can use the common semiclassical theory to study the BH model, -- we shall give an example in Sec.~\ref{E}.  

Finally, we mention that within the formalism of the Husimi function we are actually not bound with the limit $N\rightarrow\infty$ and may consider finite $N$ as well.  Of course, Eq.~(\ref{D3}) is simpler than the original Schr\"odinger equation for the many-body wave function $|\Psi(t)\rangle$ only if we neglect the last term in this equation -- the approximation known as the truncated Husimi function. Fortunately, there are physically important cases where this approximation is justified. In these cases the truncated Husimi function correctly reproduces quantum dynamics of the BH model even when the mean-field equation (\ref{D6}) fails to do this.

\subsection{Phase space of the classical Bose-Hubbard system}

Eq.~(\ref{D6}) defines classical trajectories in $2L$ dimensional phase space which lie on the energy shell $E=H({\bf a})$. Depending on the energy $E$ and initial conditions the trajectory can be either regular or chaotic. (Here we don't discuss the trivial case $L=2$ where all trajectories are regular.) It is largely an open question to find the volume of regular and chaotic components on a given energy shell. In the general case of $L$-site system we only know that there are regular islands for $E$ close to the energy of the ground. 
\footnote{More results are known for the 3-site system, where the classical phase space can be relatively easy visualized, see recent work \cite{Arwa14} and references therein.}  
As it will be shown in the next subsection, these islands are associated with the Bogoliubov spectrum for elementary excitations of a Bose-Einstein condensate (BEC). For higher energy shells, even if there are stability islands, their size is expected to be small. Statistical analysis of the energy spectrum of the quantum BH model presented in Sec.~\ref{secA} strongly supports this statement. In fact, if there were large regular islands for higher energies, we would see this as a deviation in the spectrum statistics from the Wigner-Dyson distribution. 

\subsection{Bogoliubov spectrum}
\label{E}

Let us discuss the low-energy stability islands in more detail. To make analysis simpler we shall consider $L=3$. Generalization to larger $L$ is straighforward and consists of substituting the quantity $\delta=J[1-\cos(2\pi/3)]$ in the equations below by the quantity $\delta_k=J[1-\cos(2\pi k/L)]$. 

The analysis involves several steps. First we rewrite the Hamiltonian (\ref{D4}) for $L=3$ in terms of the canonical variables $b_k$ and $b_k^*$. This gives
\begin{displaymath}
H_0=-J\sum_{k=-1}^1 \cos\left(\frac{2\pi k}{3}\right) b_k^*b_k
+\frac{g}{2}\sum_{k_1,k_2,k_3,k_4} b_{k_1}^* b_{k_2}^* b_{k_3} b_{k_4} \tilde{\delta}(k_1+k_2-k_3-k_4) \;.
\end{displaymath}
Next we switch to the action-angle variables, $b_k=\sqrt{I_k}\exp(i\phi_k)$, and explicitly take into account that $\sum_k I_k=1$. This reduces our system of three degrees of freedom to a system of two degrees of freedom:
\begin{eqnarray}
\label{E2}
H_0=(\delta+g)(I_{-1}+I_{+1}) +2gI_0\sqrt{I_{-1} I_{+1}}\cos(\phi_{-1}+\phi_{+1})  \\
\nonumber
-g(I_{-1}I_{+1}+I_{-1}^2+I_{+1}^2) +2g\sum_{\pm}I_{\mp}\sqrt{I_0 I_{\pm1}}\cos(2\phi_{\mp1}-\phi_{\pm1}) \;,
\end{eqnarray}
where $\delta=J[1-\cos(2\pi/3)]$, $I_0=1-I_{-1}-I_{+1}$ and the phases $\phi_{\pm1}$ of variables $b_{\pm1}(t)$ are measured with respect to the phase of $b_0(t)$. The low-energy dynamics of the system (\ref{E2}), which is associated with the low-energy spectrum of the quantum system, implies $I_{\pm 1}\ll I_0$. Keeping in the Hamiltonian (\ref{E2}) only the terms linear in $I_{\pm 1}$, and using one more canonical transformation,
\begin{eqnarray}
\nonumber
I=I_{+1}+I_{-1} \;,\quad \theta=(\phi_{+1}+\phi_{-1})/2 \;,\\
\nonumber
M=(I_{+1}-I_{-1})/2 \;,\quad \vartheta=\phi_{+1}-\phi_{-1} \;,
\end{eqnarray}
we obtain the effective Hamiltonian which locally describes the low-energy stability island:
\begin{equation}
\label{E4}
H_{eff}=(\delta+g)I+g\sqrt{I^2-4M^2}\cos(2\theta)  \;, \quad |M|\le I/2 \;.
\end{equation}
Note that $H_{eff}$ does not include phase $\vartheta$ and, hence, the action $M$ is a constant of motion. 
\begin{figure}[t]
\includegraphics[width=11cm, clip]{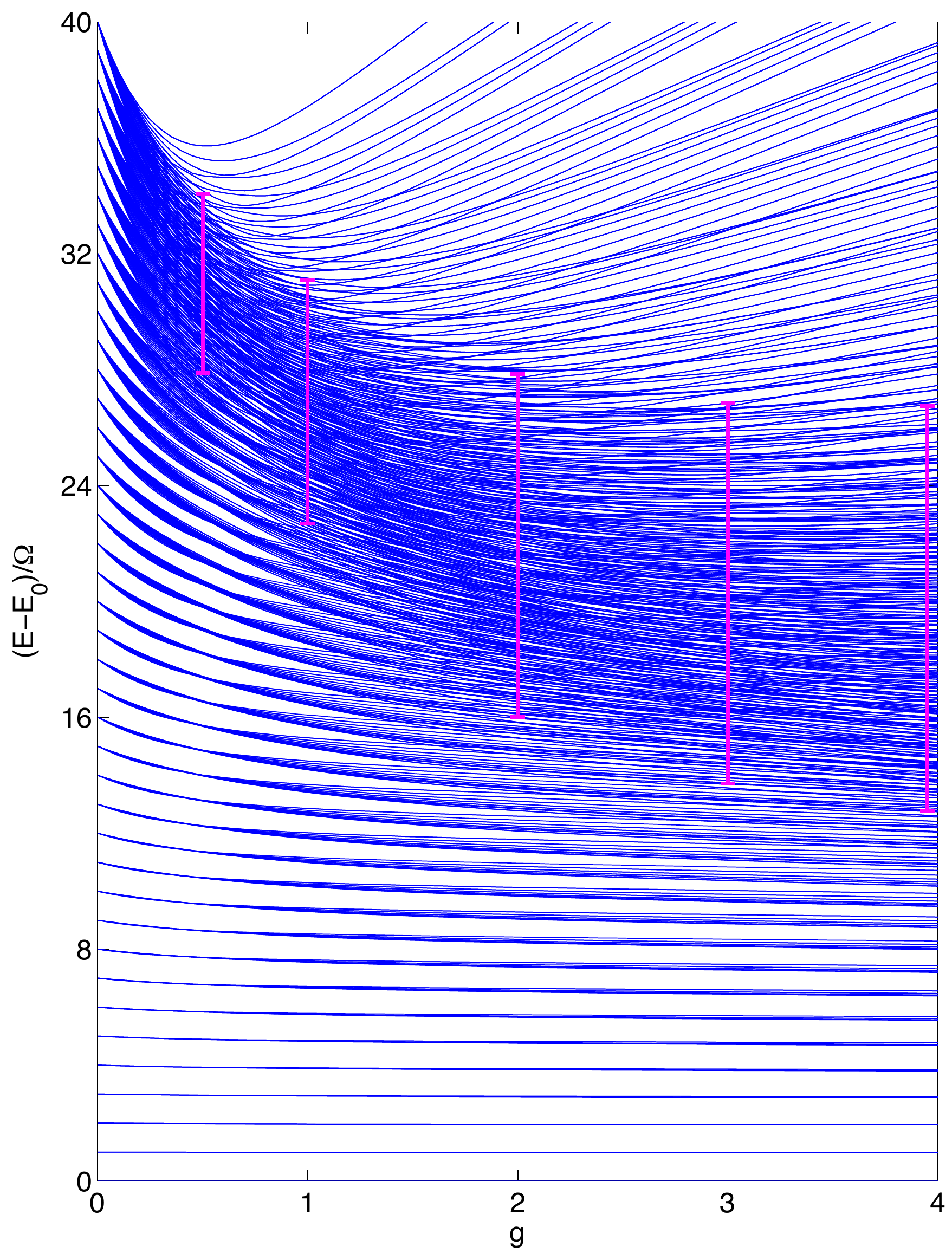}
\caption{ Energy spectrum of the 3-site BH model for $N=40$. The energy is measured with respect to the ground energy $E_0$ and scaled with respect to the Bogoliubov frequency $\Omega$. Error bars indicate energy intervals, where the classical counterpart of the system shows chaotic dynamic. The figure is borrowed from Ref.~\cite{73}.}
\label{fig7}
\end{figure}

The obtained Hamiltonian (\ref{E4}) suffices to find the low-energy spectrum of the 3-site BH model. To do this we integrate the system (\ref{E4}) by introducing new action, $\tilde{I}=(1/2\pi)\oint I(\theta,E) d\theta$, and resolving this equation with respect to the energy. We get
\begin{equation}
\label{E5}
E=\Omega \tilde{I} \;,\quad \Omega=\sqrt{2g\delta+\delta^2} \;,
\end{equation}
where the frequency $\Omega$ is nothing else as the Bogoliubov frequency. 
\footnote{For $L>3$ the system has several Bogoliubov frequencies $\Omega_k=\sqrt{2g\delta_k+\delta_k^2}\sim \sqrt{g}k$, where $k$ is usually interpreted as the wave vector of elementary excitations.}
Finally, we quantize actions $\tilde{I}$ and $M$ in units of the effective Planck constant $\hbar_{eff}=1/N$.  This gives equidistant set of energy levels $E_n=E_0+\Omega n$,  with $(n+1)$ degeneracy of the $n$th level. 

It is interesting to compare the above result with the exact energy spectrum of the 3-site model. This spectrum is shown in Fig.~\ref{fig7} where, to facilitate the comparison, we rescale it by using the Bogoliubov frequency $\Omega$. As expected, the degenerate equidistant spectrum is a good approximation only up to some critical energy, above which the classical dynamics of the BH system is chaotic. In principle, one can use the semiclassical quantization also in the chaotic region. However, this will require more sophisticated semiclassical theory known as the Periodic Orbits Theory [see Chapter 7 in Ref.~\cite{Stoe99}], which is based on the notion of the van~Vleck-Gutzwiller propagator.  An application of the van~Vleck-Gutzwiller propagator to the BH model is found in Ref.~\cite{Engl14}.

%
\section{Bloch oscillations of Bose atoms}
\label{secC}

In this section we discuss Bloch oscillations (BOs) of interacting Bose atoms. The main reason for discussing this phenomenon in the present review is that BOs can test chaotic nature of the BH model. Furthermore, recently BOs of Bose atoms in 1D lattices have been studied experimentally \cite{Mein14a}, providing the first experimental results on Quantum Chaos in the BH model.

\subsection{Governing equation}
\label{F}

In the single-band approximation (which is assumed throughout the paper) BOs of interacting atoms are described by the Hamiltonian 
\begin{equation}
\label{F1}
\widehat{H}=\widehat{H}_0 + dF\sum_l l\hat{n}_l \;,
\end{equation}
where $\widehat{H}_0$ is the BH Hamiltonian (\ref{A1a}), $F$ a static (for example, gravitational) field, and $d$ the lattice period ($d=1$ in what follows). Introducing the Bloch frequency $\omega_B=dF/\hbar\equiv F$ and using the substitution $\hat{a}_l \rightarrow \hat{a}_l\exp(-i F l t)$ the time-independent Hamiltonian (\ref{F1}) reduces to the form (\ref{C5}) where $\theta=F t$:
\begin{equation}
\label{F2}
\widehat{H}(t)=-\frac{J}{2} \sum_{l=1}^L \left( \hat{a}^\dag_{l+1}\hat{a}_l e^{iFt}+h.c.\right)
  +\frac{U}{2}\sum_{l=1}^L \hat{n}_l(\hat{n}_l-1)  \;.
\end{equation}
Referring to the experiment \cite{Mein14a} the initial wave function of interacting atoms is given by the ground state of the Hamiltonian $\widehat{H}_0=\widehat{H}(t=0)$ and the simplest quantity to be measured is the mean momentum per one atom:
\begin{displaymath}
p(t)=-\frac{J}{N}{\rm Im}\langle \Psi(t)|\sum_l \hat{a}^\dagger_{l+1}\hat{a}_l e^{iFt}|\Psi(t)\rangle \;. 
\end{displaymath}
Notice that for non-interacting atoms we would have $p(t) = -J\sin(F t)$. This equation is the essence of phenomenon of BOs. The problem to be addressed is the effect of atom-atom interactions which, as it was shown in Sec.~\ref{secA}, make the Hamiltonian (\ref{F2}) chaotic.

\subsection{Mean-field analysis}

Let us first analyze BOs of interacting atoms by using the mean-field approach:
\begin{equation}
\label{F4}
i \frac{d a_l}{dt}=-\frac{J}{2}\left(a_{l+1} e^{iF t} +a_{l-1}e^{-iF t}\right) + g|a_l|^2 a_l  \;,
\end{equation}
It is easy to check that Eq.~(\ref{F4}) has a periodic solution
\footnote{Not periodic phase $\exp(-igt)$ in Eq.~(\ref{F5}) is irrelevant and can be removed by the obvious substitution.} 
%
\begin{equation}
\label{F5}
a_l(t)=\exp\left(i\frac{J}{F}\sin(F t)-igt\right) \;.
\end{equation}
For the mean momentum per atom Eq.~(\ref{F5}) gives 
\begin{displaymath}
p(t)=-\frac{J}{L}{\rm Im}\left(\sum_l a^*_{l+1}a_l e^{iFt} \right)= -J\sin(F t) \;.
\end{displaymath}
Thus one can consider the solution (\ref{F5}) as a candidate for BOs of interacting atoms, where the next step is stability analysis of the periodic trajectory (\ref{F5}).

\begin{figure}
\includegraphics[width=8cm, clip]{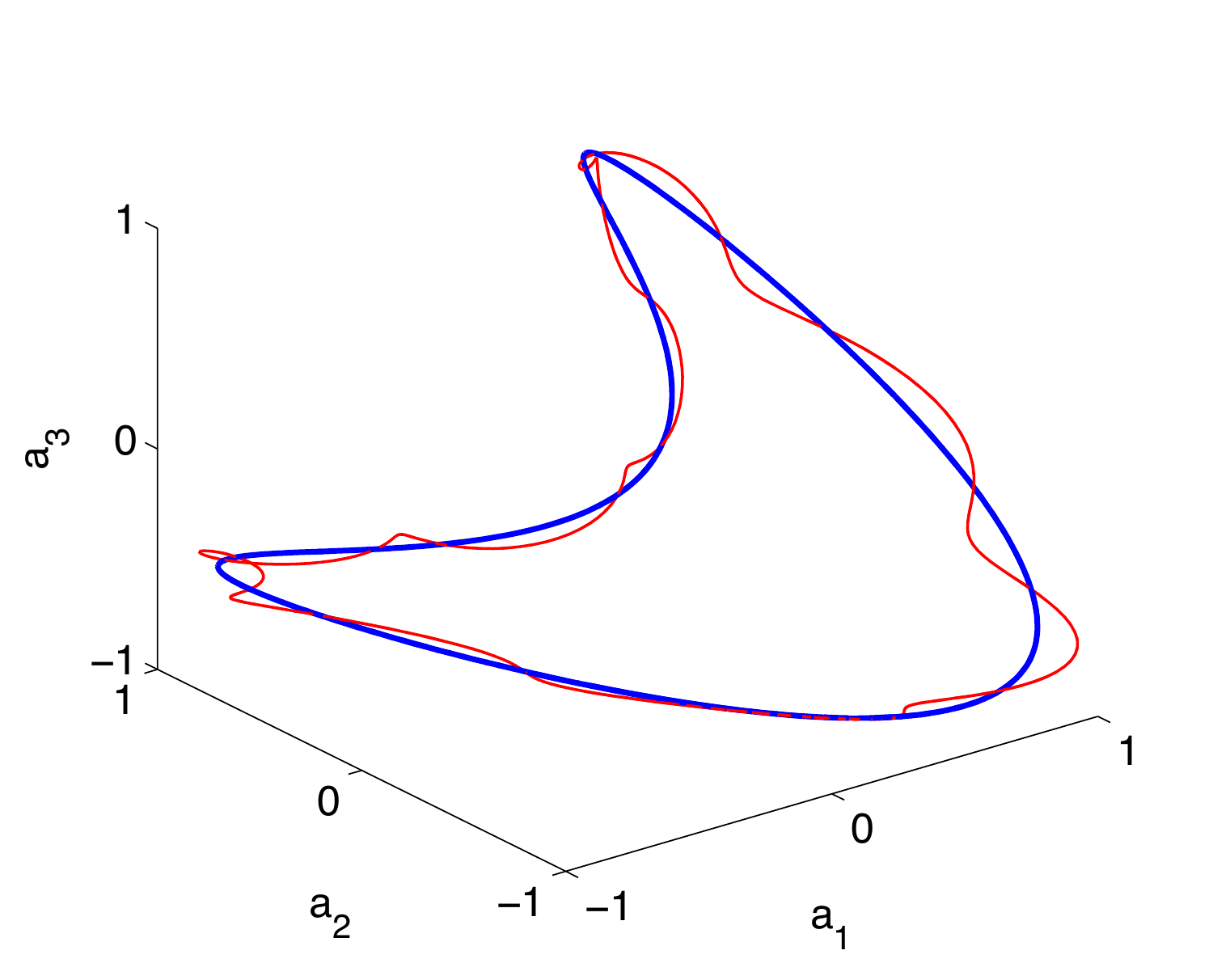}
\hspace{0.5cm}
\includegraphics[width=8cm, clip]{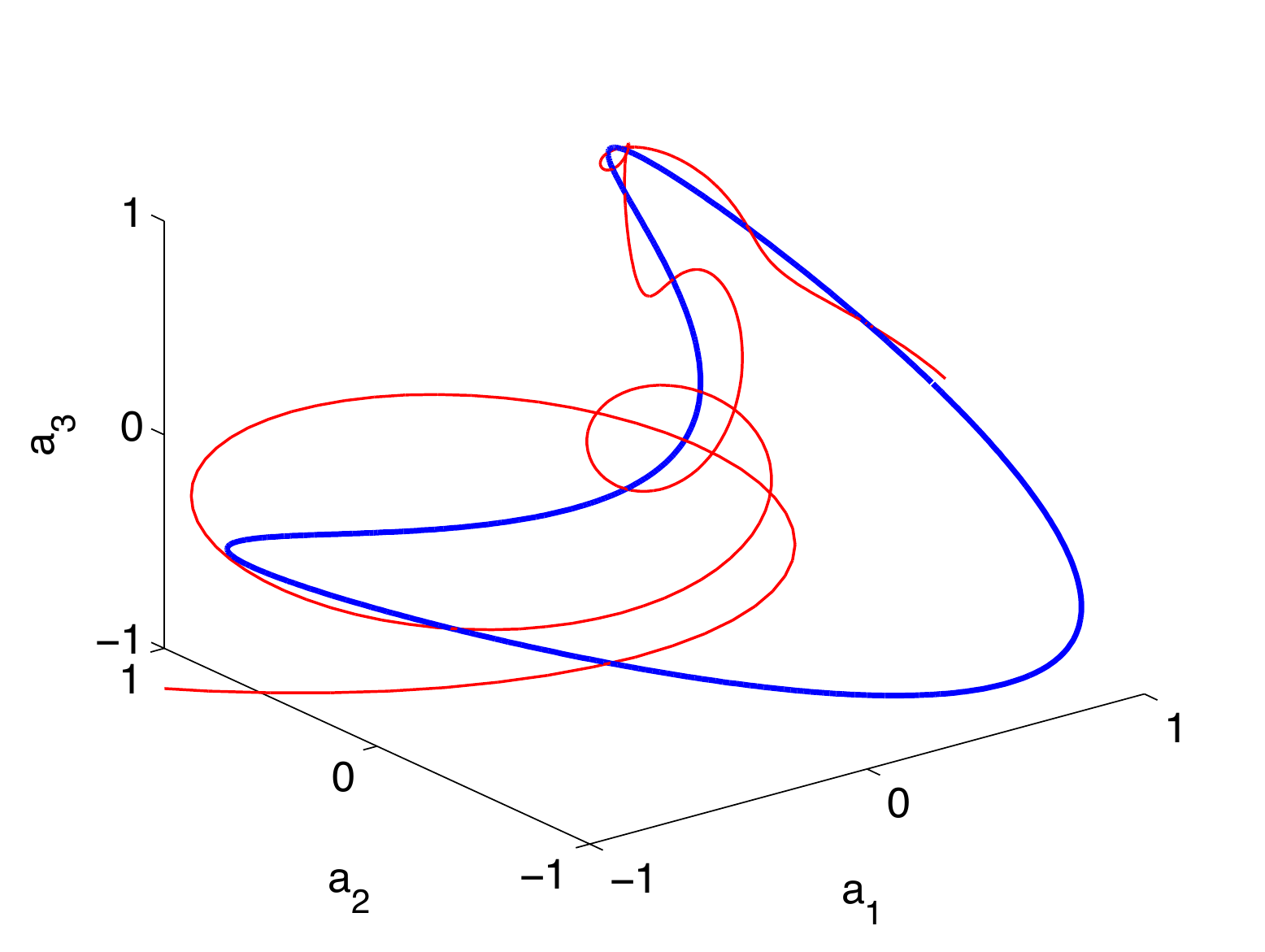}
\caption{ Pictorial presentation of the stable and unstable periodic trajectory in the multi-dimensional phase-space of the classical BH system.}
\label{fig8}
\end{figure}

The stability analysis is done in the usual way, i.e., by linearizing Eq.~(\ref{F4}) around the periodic trajectory (\ref{F5}): 
\begin{equation}
\label{F3}
i\frac{d}{dt}\delta {\bf a}={\cal M}[{\bf a}(t)]\delta {\bf a}  
\end{equation}
(here $\delta {\bf a}(t)$ is a deviation from the periodic trajectory and ${\cal M}$ the Jacobi matrix). It leads to the following result  \cite{Zhen04,80}. In the limit of large $L$ the parameter space of the system (\ref{F4}) is divided into two parts by the critical line
\begin{equation}
F_{cr}\approx\left\{
\begin{array}{ll}
3g \;, & F<2J \\ \sqrt{10gJ} \;, & F>2J
\end{array}\right. \;.
\end{equation}
In the strong field regime $F>F_{cr}$ all Lyapunov exponents of the linear Eq.~(\ref{F3}) are zero and, hence, the solution (\ref{F5})  is stable. In the opposite case $F<F_{cr}$ there are positive exponents and the solution is unstable. Pictorially, this result is illustrated in Fig.~\ref{fig8} which shows periodic trajectory in the multi-dimensional phase-space of the system together with a nearby trajectory. 

\subsection{Quantum ensemble}

\begin{figure}
\includegraphics[width=11cm, clip]{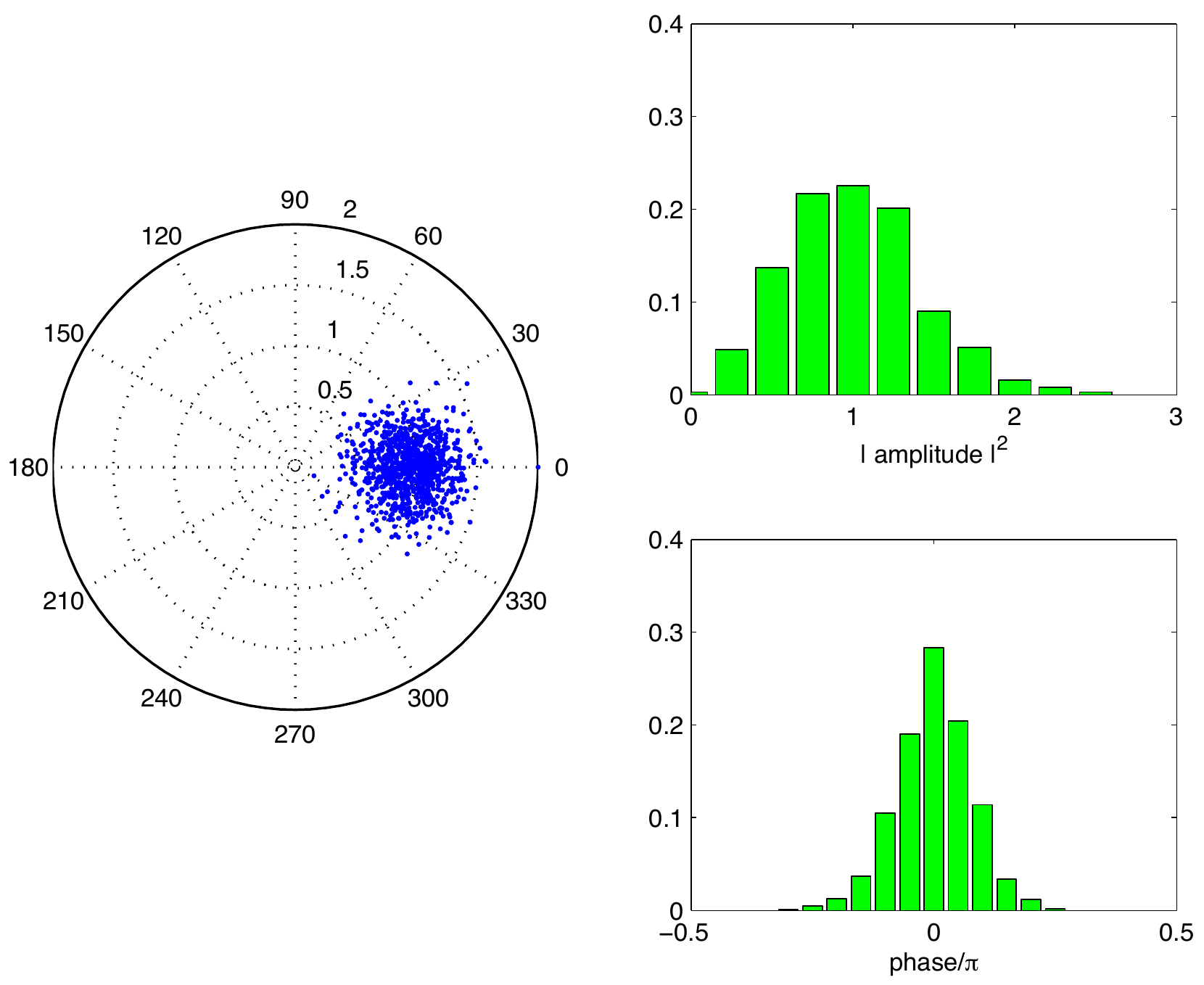}
\caption{Quantum ensemble (100 realizations) representing the many-body BEC state for $N=15$ and $L=5$. The characteristic widths of distributions are inverse proportional to $\sqrt{\bar{n}}$.}
\label{fig9}
\end{figure}

We have shown that the trajectory (\ref{F5}), which is defined by the initial conditions $a_l(t=0)=1$,  is unstable in the weak field regime. Thus an arbitrary small deviation from the specified initial conditions results in a completely different trajectory. 
\footnote{This phenomenon is often referred to as the dynamical or modulation instability.}
This brings us back to the Husimi function, which describes the evolution of not a single trajectory but of an ensemble of trajectories. This ensemble is obviously defined by the equation 
\begin{equation}
\label{F6}
f({\bf a},t=0)=|\langle {\bf a} | \Psi(t=0)\rangle |^2 \;,
\end{equation}
where $|\Psi(t=0)\rangle$ is the initial many-body wave function given by the ground state of the Hamiltonian $\widehat{H}_0$. For $U< J$ and $\bar{n}\sim 1$ this ground state is well approximated by a BEC of non-interacting atoms, 
\begin{equation}
\label{F7}
|\Psi(t=0)\rangle = \left(\hat{b}^\dagger_{k=0}\right)^N | vac \rangle \;, \quad
\hat{b}^\dagger_{k=0}=L^{-1/2}\sum_l \hat{a}_l^\dagger \;.
\end{equation}
Then the distribution (\ref{F6}) is known analytically and we can generate ensemble of initial conditions by using, for example, the exception-rejection numerical method \cite{Pres07}. In what follows we shall refer to this ensemble of initial conditions as the quantum ensemble, to stress that it is defined by the quantum many-body state of the system. As an example, Fig.~\ref{fig9} shows the quantum ensemble for the state (\ref{F7}) in the Wannier representations for $L=5$ and $N=15$. It is seen that amplitudes $a_l(t=0)$ deviates from unity by both the absolute value and the phase, where the characteristic size of deviations is inverse proportional to $\sqrt{\bar{n}}$.

\begin{figure}
\includegraphics[width=8.5cm, clip]{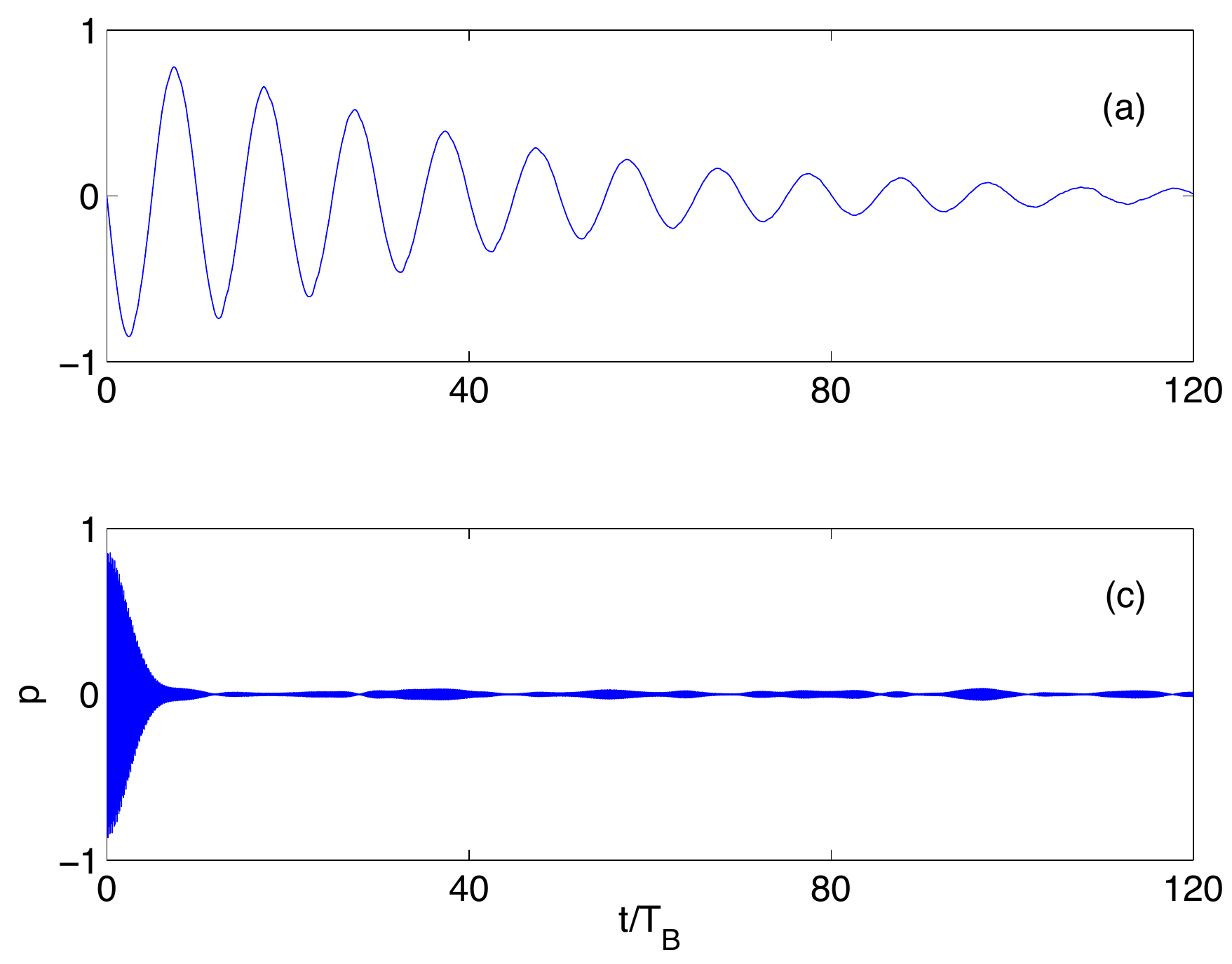}
\hspace{0.5cm}
\includegraphics[width=8.5cm, clip]{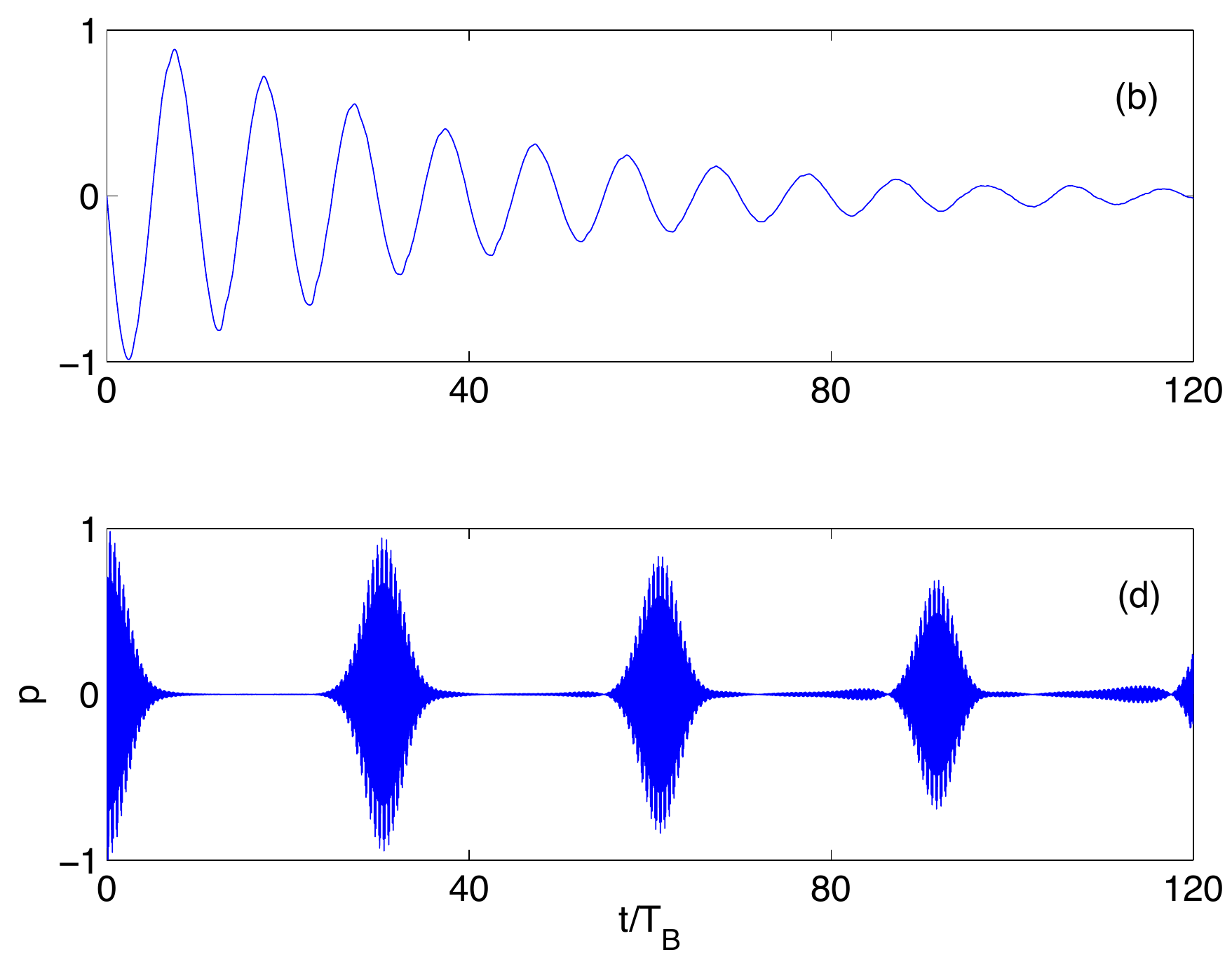}
\caption{Bloch oscillations of interacting atoms calculated by using the truncated Husimi function (left) and the full quantum-mechanical simulations (right). The system parameters are $N=15$, $L=5$, $J=1$, $U=0.1/3$, and $F=0.1$ (top), and $F=10$ (bottom).}
\label{fig10}
\end{figure}

Having the quantum ensemble in hands we evolve each trajectory according to Eq.~(\ref{F4}) and average the result over the ensemble. (Clearly, this amounts to the Monte-Carlo solution of the truncated equation on the Husimi function.)  The panel (a) in Fig.~\ref{fig10} shows the mean atomic momentum as function of time for the static field $F<F_{cr}$. An exponential decay of BOs,
\begin{equation}
\label{G1}
p(t)=-J\exp(-\gamma t)\sin(F t) \;,
\end{equation}
is clearly seen. The panel (a) should be compared with the panel (b) showing the solution of the Schr\"odinger equation with the microscopic Hamiltonian (\ref{F2}).  An excellent agreement is noticed. We discuss the physics behind this remarkable result  in the next subsection.

\subsection{Internal decoherence}
\label{G}

Previous studies of the Bloch dynamics of interacting Bose atoms \cite{61,80} proved that exponential decay of BOs is due to decoherence of the initial BEC state or self-thermalization of the system. On the formal level the self-thermalization  means that the one-particle density matrix ${\cal R}(t)$ relaxes to a diagonal (in the Bloch basis) matrix with equal populations of the single-particle quasimomentum modes. Thus the linear entropy of the system $S={\rm Tr}({\cal R}^2)$, which is one of possible characteristics of the system coherence, decreases from unity to $S=1/L\ll 1$. From the viewpoint of classical mechanics the self-thermalization is a consequence of chaotic dynamics of the system or, more precisely, the mixing property of the chaotic dynamics.

The discussed `internal' decoherence has common features with `external' decoherence caused by interaction of the system with the environment.  In particular, if we consider BOs of a single atom coupled to a bath, 
\footnote{In the problem of atomic BOs the bath is given by zero modes of the electromagnetic field which are responsible for spontaneous emission. In laboratory experiments intensity of this process and, hence, the rate of external decoherence is controlled by tuning the laser frequency further or closer to the atomic resonance.} 
we shall also observe the exponential decay of oscillations \cite{56}. On the other hand, we know from studies on the general problem of quantum-classical correspondence that external decoherence suppresses interference terms in the governing equation of motion \cite{Zure91,23,28}.  In Eq.~(\ref{D3}) on the Husimi function these terms are denoted as $O(1/N)$. The excellent agreement between exact quantum simulations and the semiclassical approach of  the truncated Husimi function proves that this is also the case for the internal decoherence. Thus we have a loop: classical chaotic dynamics is responsible for the internal decoherence which, in its turn, causes the quantum system to behave classically.

It is interesting to consider the strong field limit $F>F_{cr}$ which breaks the above loop. Now the classical dynamics is stable and, hence, there is no internal decoherence. Dynamics of the mean momentum for $F>F_{cr}$ is shown in the panel (d) in Fig.~\ref{fig10}. It presents periodic revivals of BOs which are described by the following simple equation \cite{57},
\begin{equation}
\label{G2}
p(t)=-J\exp\left(-2\bar{n}[1-\cos(Ut)]\right)\sin(F t) \;,\quad  \bar{n}=N/L \;.
\end{equation}
It  is seen in Fig.~\ref{fig10}(c) that the truncated Husimi function approach does not reproduce the revivals, which are a quantum interference effect. 
Recent laboratory studies of BOs in 1D lattices \cite{Mein14a} undoubtedly confirm the transition from  the quasi-periodic dynamics (\ref{G2}) to the exponentially decaying oscillations (\ref{G1}) as the static field $F$ is decreased.
\footnote{In the cited experiment the authors used gravitational field which was compensated by the levitation force to a desired level.}

\section{Conclusions}

We discussed the Bose-Hubbard Hamiltonian beyond its ground state. It was proven that this system is a chaotic system in the sense of Quantum Chaos. Of course, it is not the only many-body system which is chaotic -- the other examples are provided, for instance, by non-integrable models of spin chains \cite{Mont93,Berm01}. However, the BH model is extremely important in the experimental cold-atom physics and for this reason deserves a special attention. 

In the review we focused on the case where the kinetic energy of atoms dominates the interaction energy ($J>U$).  In this case the ground state of the system is known to be a super-fluid state with low-energy excitations described by the Bogoliubov theory. We revisited this problem from uncommon perspective of the quantum-classical correspondence and showed that the Bogoliubov spectrum can be obtained by quantizing the low-energy stability islands of the classical BH model.

If we go to higher energies, the energy spectrum of the BH model become irregular and must be analyzed statistically. We considered the simplest statistical characteristic of the spectrum -- the level spacing distribution -- which was shown to obey the Wigner-Dyson equation for random matrices. 

Another direction of research is response of the BH system to a sudden change of its parameters (so-called quench dynamics) or to external perturbations. In the review we considered the response to a static field. For cold atoms in optical lattices this could be, for example,  the gravitational field \cite{Poli11}. For the initial condition given by the ground state of the system, the static field induces Bloch oscillations which, according to the mean-field analysis, can be either stable or unstable. The microscopic analysis of the system shows that the unstable regime results in the exponential decay of BOs due to internal decoherence, which is a consequence of the chaotic dynamics of the system. Remarkably, in this case we were able to reproduce the quantum dynamics by using the `classical' approach of the truncated Husimi function. This result  sheds the new light on the old problem of the onset of classicality in our world.

We conclude this review by mentioning some future prospects. As stated above, we focussed on the case $J> U$ where the ground state of the BH model is a super-fluid state. It is interesting to study the opposite case $U\gg J$ (where the ground state of the system is a Mott insulator) by using a semiclassical approach. In particular, this concerns the problem of quantum phase transition from the Mott-insulator state to the density-wave state in tilted 1D lattices \cite{Sach02,Simo11}. The second avenue is a generalizing the results to two-dimensional case where, besides potential fields, one can include into consideration gauge fields \cite{Aide11,Aide13,Miya13}. Since atoms are neutral these fields are often referred to as synthetic magnetic fields. 
%
For vanishing interactions dynamical and spectral properties of cold atoms in a 2D lattice subject to the synthetic magnetic and electric fields are discussed in detail in the recent work \cite{preprint}.


\end{document}